\documentclass[fleqn,usenatbib]{mnras}

\usepackage{newtxtext,newtxmath}

\usepackage[T1]{fontenc}
\usepackage{ae,aecompl}

\usepackage{dsfont}
\usepackage{subfigure}
\usepackage[utf8]{inputenc}
\usepackage{pgfplots}

\usepackage{graphicx}	
\usepackage{amsmath}	
\usepackage{amssymb}	

\usepackage{color}
\usepackage{soul}
\usepackage{amssymb}

\title[Evolving grain-size distributions]{Evolving grain-size distributions embedded in gas flows}

\author[R. Sumpter \& S. Van Loo]{
  R. Sumpter$^{1}$\thanks{E-mail: py12rs@leeds.ac.uk}
  and S. Van Loo$^{1}$\\
  $^{1}$School of Physics and Astronomy, University of Leeds, Woodhouse Lane, Leeds LS2 9JT, UK 
}

\date{Accepted XXX. Received YYY; in original form ZZZ}

\pubyear{2019}

\begin{document}
\label{firstpage}
\pagerange{\pageref{firstpage}--\pageref{lastpage}}
\maketitle

\begin{abstract}
We present a numerical approach for accurately evolving a dust grain-size distribution undergoing number-conserving (such as sputtering) and/or mass-conserving (such as shattering) processes.
As typically observed interstellar dust distributions follow a power-law, our method adopts a power-law discretisation and uses both the grain mass and number densities in each bin to determine the power-law parameters. This power-law method is complementary to piecewise-constant and linear methods in the literature.
We find that the power-law method surpasses the other two approaches, especially for small bin numbers. In the sputtering tests the relative error in the total grain mass remains below 0.01\% independent of the number of bins $N$, while the other methods only achieve this for $N > 50$ or higher. 
Likewise, the shattering test shows that the method also produces small relative errors in the total grain numbers while conserving mass. Not only does the power-law method conserve the global distribution properties, it also preserves the inter-bin characteristics so that the shape of the distribution is recovered to a high degree. This does not always happen for the constant and linear methods, especially not for small bin numbers.
Implementing the power-law method in a hydrodynamical code thus minimises the numerical cost whilst maintaining high accuracy. The method is not limited to dust grain distributions, but can also be applied to the evolution of any distribution function, such as a cosmic-ray distribution affected by synchrotron radiation or inverse-Compton scattering.
\end{abstract}




\section{Introduction}
Within the Interstellar Medium (ISM), dust grains are an important ingredient as they lock up substantial 
fractions of the heavy elements \citep{Draineetal2007}, produce the dominant contribution to the opacity 
for radiation upward of the Lyman limit \citep{DraineLee1984}, contribute a significant part of gas heating 
through
photoelectric heating \citep{BakesTielens1994} and provide the surface onto which chemical elements can 
accrete and react \citep{GarrodHerbst2006}. They also make up a significant fraction of the mass of the 
ISM with a canonical value of about 1\%. This value comes from observational constraints by fitting extinction,
scattering or polarisation of background stellar radiation and IR dust emission \citep[e.g.][]{knapp74, jura79}. These restrictions further imply that the dust grains follow a power-law size distribution given by 
\begin{equation}\label{eq:MRN}
	 \frac{dn(a)}{da} \propto a^{-3.5},
\end{equation}
where $a$ is the grain radius and $\displaystyle \frac{d n(a)}{da} da$ is the number density of grains with radii in the range $[a, a+da]$ \citep[][]{MRN77,WeingartnerDraine2001}.

Although the typical grain-size distribution follows the \citet{MRN77} (MRN) distribution, local variations
are expected as the grain-size distribution is set by balancing dust production, growth and destruction processes. 
It is thought that dust grains are produced in the upper atmospheres of asymptotic giant branch (AGB) stars 
\citep{Maercker18} as well as formed \citep{nozawa03} and destroyed \citep{kirchschlager19} 
in supernova remnants. In the dense, quiescent regions of molecular clouds grains primarily grow in size due 
to coagulation and mantle accretion \citep{jones85,liffman89,ossenkopf93,inoue03,ormel09,asano13,ysard16,jones17}. In 
contrast, in star forming regions, shocks mainly accommodate the destruction of dust grain sizes due to sputtering, 
shattering and vaporisation \citep{tielens94, jones96, flower03, hirashita09, guillet07, guillet09, guillet11, 
anderl13, vanloo13}. In protoplanetary discs both growth and fragmentation of dust grains take place
\citep{brauer08, birnstiel10, birnstiel18, dullemond18, homma18, tamfal18}.

As these grain processes affect the overall grain distribution, they can have significant 
effects on the dynamics of e.g. the ISM and protoplanetary discs. For example, in the outflows of Young
Stellar Objects (YSOs), dust grains are important charge and current carriers and therefore determine the 
structure of C-type shocks \citep{VanLooetal2009}, while, in protoplanetary discs, different dust grain
sizes influence the growth and structure of the magnetorotational instability \citep{SalmeronWardle2008}.
Furthermore, grain-processing leads to observational signatures. In typical ISM conditions silicon is 
adsorbed onto dust grains. However, gas phase SiO is detected in the clumpy structure of YSO outflows
due to the shock-induced sputtering releasing silicon into the gas phase \citep[e.g.][]{Martin-Pintadoetal1992,
Mikamietal1992}. Thus, it is essential to accurately follow the evolution of the dust grain distribution 
to model both the dynamics and emission signatures.

Previous studies of the dust processing have used different methods. In the simplest approach only a few
dust species with specified radii, i.e. typically, one or two grain species representing small and/or 
large dust grains, are evolved \citep[e.g.][]{DRD83, vanloo13, Hirashita2015}.  While this is appropriate 
to capture number-conserving grain processes such as sputtering and mantle accretion, it does not 
adequately model mass-conserving processes such as shattering\footnote{We note that shattering is not 
always strictly mass-conserving, since fragments may be produced which are smaller than the minimum grain 
radius limit of the distribution.} and coagulation. A more rigorous approach is 
to follow the dust grain distribution using a discrete grain-size distribution. Here the distribution is updated, 
by considering number conservation or mass conservation, to
redistribute the grains across the bins in a way appropriate to capture the modelled grain process 
\citep[e.g.][]{mizuno88,jones94, jones96,liffman89,mckinnon18}. As most of these studies focus on coagulation
and shattering, i.e. the mass-conserving grain processes,
these models only follow the total dust mass and not the total grain number. In order to also model sputtering and mantle accretion, \citet[][hereafter McK18]{mckinnon18} modified the discrete distribution approach by using a piecewise-linear discretisation in each grain-size bin to conserve both mass and particle numbers. They show that this technique is second-order accurate in the number of size bins. However, to achieve an accuracy of the order 
of 1\% in both mass and number conservation, it is necessary to use about 50 to 100 bins. Thus, including such 
a scheme into a numerical hydrodynamics (HD) or magnetohydrodynamics (MHD) code significantly increases
the computational cost. 

In this paper we propose a different approach to the discretisation within a size bin in order to 
minimise the numerical cost while maintaining high accuracy. As the typical observed dust distribution 
follows a power-law, i.e. the MRN distribution, we will adopt a power-law discretisation. 
Section~\ref{sec:method} outlines the method describing the determination of the power-law 
coefficient and index, and the redistribution of the dust grains for both number-conserving and 
mass-conserving processes. In Section~\ref{sec:results} the method is applied to a number of test problems
and the results are discussed. Then Section~\ref{sec:equations} describes the modifications needed to the HD 
and MHD equations in order that these reflect the power-law distribution of the dust grains, and, finally, 
our conclusions are given in Section~\ref{sec:conclusions}.

\section{Numerical method for grain distribution evolution}\label{sec:method}
Here we will describe the numerical methods to evolve a grain power-law distribution. First we 
consider the construction of a piecewise power-law distribution and the formulation of  the 
power-law coefficients and indices. As we compare the power-law method to piecewise-linear and 
piecewise-constant discretisations in Sect.~\ref{sec:results}, we also present these formulations and 
explain how they relate to one another. Finally, the routines to redistribute mass and number 
density of grains across the distribution bins are discussed for grain processes that conserve the grain 
numbers or  total grain mass.

\subsection{Discrete power-law distribution}\label{sec:discrete}
Although dust grains are generally irregularly shaped \citep[e.g.][]{Draine2003}, for the purpose of this 
paper we assume they are spherical. This significantly simplifies the treatment of the dust grains as 
the grain distribution depends only on the grain radius $a$. Furthermore, we assume that the range of grain
radii is limited to the range $[a_{\rm min}, a_{\rm max}]$. This range is then divided up logarithmically 
with a spacing determined by 
\begin{equation}\label{eq:delta}
	\Delta a = \frac{\log\left(a_{\rm max}/a_{\rm min}\right)}{N},
\end{equation}
where $N$ is the number of bins. This means that the edges of bin $i$ are effectively
\begin{equation}\label{eq:ai}
	\begin{aligned}
		&a_{i} = a_{\rm min}\mathrm{e}^{i \Delta a} \\
		&a_{i+1} = a_{\rm min}\mathrm{e}^{(i+1)\Delta a}
	\end{aligned}
\end{equation}
where $i = 0, 1, ..., N-1$.
Now we assume that the differential grain-size number density distribution in bin $i$ has a 
power-law shape, i.e.
\begin{equation}\label{eq:sizeDist}
\left.\frac{\partial{n(a, t)}}{\partial{a}}\right|_i = A_i a^{-\alpha_{i}}, 
\end{equation}
where $\displaystyle \frac{\partial{n(a,t)}}{\partial{a}} da$ is the number density of grains in a size interval 
$[a, a+da]$ at time $t$. Note that, as grain processes change the distribution function, the power-law 
coefficient $A_i$ and index $\alpha_i$ are implicitly time-dependent.

To determine the power-law coefficient and index we use the grains' bin-averaged number density, $n(t)$, and mass 
density, $\rho(t)$, which are followed according to the redistribution routine in Sect.~\ref{subsec:Redist}. The 
bin-averaged number density in bin $i$ at time $t$ is given by the integral  
\begin{equation}
	n_i(t) = \int\limits_{a_i}^{a_{i+1}}{\left.\frac{\partial{n(a, t)}}{\partial{a}}\right|_i da} = \left\{
		\begin{array}{ll}
			\frac{A_i}{1 - \alpha_i}\left(a^{1-\alpha_i}_{i+1} - a^{1-\alpha_i}_{i}\right) & \alpha_i \neq 1\\
			\\
			A_i \log\left(a_{i+1}/a_i\right) & \alpha_i = 1 
		\end{array}\right.
\end{equation}
It is actually convenient to use Eq.~\ref{eq:ai} to rewrite this expression as 
\begin{equation}\label{eq:ni}
	n_i(t) = A_i a^{1 - \alpha_i}_{i+1/2} \Delta a\ \mathcal{F}\left((1 - \alpha_i) \frac{\Delta a}{2}\right),
\end{equation}
where $\mathcal{F}(x) = \sinh(x)/x$ and applies to all values of $\alpha_i$, and
	$a_{i+1/2} = a_{\rm min}\mathrm{e}^{(i+1/2)\Delta a}$.
However, this does not uniquely determine $A_i$ and $\alpha_i$. Therefore we need a second expression 
provided by the grains' bin-averaged mass density
\begin{equation}
	\rho_i(t) = \int\limits_{a_i}^{a_{i+1}}{m(a)\left.\frac{\partial{n(a, t)}}{\partial{a}}\right|_i da},
\end{equation}
where 
\begin{equation}\label{eq:m}
m(a) = \frac{4\pi \rho_g}{3} a^3
\end{equation}
is the grain mass with $\rho_g$ the density of the grain material.
Similarly to Eq.~\ref{eq:ni} we obtain
\begin{equation}\label{eq:rhoi}
	\rho_i(t) = \frac{4\pi \rho_g}{3} A_i a^{4 - \alpha_i}_{i+1/2} \Delta a\ \mathcal{F}\left((4 - \alpha_i) \frac{\Delta a}{2}\right).
\end{equation}
By combining Eqs.~\ref{eq:ni} and \ref{eq:rhoi} and recognising that the average grain mass in bin $i$ is 
$m_i = \rho_i/n_i$, we derive an expression that solely depends on the power-law
index $\alpha_i$, i.e.
\begin{equation}\label{eq:root}
	\frac{m_i}{m(a_{i+1/2})}  \mathcal{F}\left((1 - \alpha_i) \frac{\Delta a}{2}\right)
	- \mathcal{F}\left((4 - \alpha_i) \frac{\Delta a}{2}\right) = 0
\end{equation}
This expression needs to be solved numerically using a root-finding algorithm, for example the Newton-Raphson method. As 
$\mathcal{F}$ in Eq.~\ref{eq:root} is a monotonic function, only a few iterations are needed to find the 
solution, especially if the initial guess is close to the root. Once $\alpha_i$ is determined, the value of 
$A_i$ can be directly calculated from Eq.\ref{eq:ni} or Eq.~\ref{eq:rhoi}.

The power-law description can be compared to methods previously used in the literature such as the 
piecewise-constant and piecewise-linear ones \citep[e.g.][McK18]{mizuno88,jones94,jones96,liffman89}.
The piecewise-constant discretisation takes on a constant value for the distribution in bin $i$ according to
\begin{equation}
       \left.\frac{\partial{n(a, t)}}{\partial{a}}\right|_i = \frac{n_i(t)}{(a_{i+1} - a_{i})},
\end{equation}
where we have chosen here that the constant reflects the total number density of grains in the bin. 
Alternatively, one can choose the distribution to reflect the mass density of the grains in the bin. A clear
disadvantage is that this method only accurately describes the number density or the mass density, but not both.
The piecewise-linear method of McK18 fixes this by assuming a linear distribution around the bin's midpoint
$a_{c,i} = (a_i + a_{i+1})/2$
\begin{equation}\label{eq:lin}
  \left.\frac{\partial{n(a, t)}}{\partial{a}}\right|_i = \frac{n_i(t)}{(a_{i+1} - a_{i})} + s_{i}(t)(a - a_{c,i}),
\end{equation}
where the slope $s_i(t)$ is chosen so that the mass density in the bin is equal to $\rho_i$. Note, however,
that the linear distribution can become negative and non-physical if the slope is too steep. This is remedied 
by imposing a slope limiter ensuring positivity of the distribution function and conserving grain mass 
density (see Section 3.2.1 of McK18). Unfortunately, this also implies that the grain numbers within the bin 
are not conserved. The piecewise-constant and piecewise-linear methods can be considered to be first and 
second-order approximations to the power-law, respectively. The accuracy depends on the bin size and on
the distribution that needs to be modelled. For example, if the distribution is flat within the bin,
all three methods give identical results as $s_i = 0$ in the piecewise-linear method and $\alpha_i = 0$ in the 
power-law method. 

In describing the methods we have implicitly assumed that the grain distribution fills an entire bin. 
This does not necessarily need to be true, especially near the limits of the distribution $r_{\rm min}$ and 
$r_{\rm max}$
(where $a_{\rm min} \leq r_{\rm min} < r_{\rm max} \leq a_{\rm max}$). Then it is possible that the 
distribution is skewed towards one of the bin edges. In the piecewise-linear method of McK18, this causes
the distribution function to become negative within the bin and slope limiting is required.
On the other hand, the power-law method always guarantees positivity, but, unfortunately, a skewed distribution produces a large power-law index resulting, numerically, in a floating point error. 
To take into account the possibility that bin~$i$ is only partly populated, we set $a^*_i = \max(a_i, r_{\rm min})$ and $a^*_{i+1} = \min(a_{i+1}, r_{\rm max})$ as bin limits. Furthermore, we need to take $\Delta a^* = \log\left(a^*_{i+1}/a^*_i\right)$ and $a^*_{i+1/2} = a^*_i e^{\Delta a^*/2}$ to determine $A_i$ and $\alpha_i$. This small modification avoids floating point errors and conserves both grain mass and numbers. Note that this also means we are not restricted to logarithmically uniform bin widths, but can have randomly sized bins. 

While using $\Delta a^*$ in the root finding algorithm for Eq.~\ref{eq:root} improves the conservation properties of
the distribution function, it also highlights a concern when $\Delta a^*$ becomes small, i.e. the root finding 
algorithm does not find a unique solution for $\alpha_i$, or finds no solution at all. This is due to the shape of 
the function $\mathcal{F}$. For small values of $x$, and thus $\Delta a^*$, the function reduces to $\mathcal{F}(x) \approx 1 + \frac{x^2}{3\!}$ and is numerically a constant for $x < 1.5 \times 10^{-3}$ (as the second term falls below the machine precision of $10^{-7}$). Given that $\Delta a^*$ is of the same order as $x$, we are limited to using the root finding algorithm for values of $\Delta a^* > 5\times 10^{-3}$.  For values below this limit, we opt to use $\alpha_i = 0$ and thus assume that the distribution is constant within the bin.

\subsection{Redistribution of grain numbers and mass} \label{subsec:Redist}
The evolution of an advected dust grain-size distribution can be expressed as \citep{TsaiMathews1995}
\begin{equation}\label{eq:advection}
	\frac{\partial}{\partial t}\left(\frac{\partial n(a,t)}{\partial a}\right) + 
	\nabla \cdot \left(\frac{\partial n(a,t)}{\partial a} \bf{v}\right) =  
	- \frac{\partial}{\partial a}\left(\frac{\partial n(a,t)}{\partial a} \frac{da}{dt}\right) 
	+ S(a, t),
\end{equation}
where $da/dt$ is the rate of change of the grain radius and $S(a,t)$ a source or sink of grains. Note that, 
if the right-hand side terms are equal to zero, this just represents changes in the distribution due to 
advection. Therefore, physical processes that affect the grain-size distribution are described by the terms 
on the right-hand side of the expression. The first term represents processes that increase or decrease the 
grain radius and conserve the total grain numbers, e.g. sputtering and mantle accretion, while 
production and destruction processes are included in the second one. Of the latter processes we only focus on 
the ones that conserve mass, such as shattering and coagulation, as other processes like supernova dust 
production are straightforward to implement. In the following we use sputtering and shattering, which are 
relevant to C-type shocks, to illustrate the methods for the grain distribution evolution. 

\subsubsection{Number-conserving processes}\label{sect:numbcons}
In the ISM the impact of neutral particles and ions on dust grains releases, or sputters, grain material 
such as Si, Mg, and O, at a rate \citep{tielens94}
\begin{equation}
	\frac{d N_s}{dt} = 2\pi a^2 n_p u_p Y_{\rm s}(u_p),
	\end{equation}
where $N_s$ is the number of sputtered species, $n_p$  the number density of impact particles, $u_p$ the relative speed between 
the impacting particles and the grains and $Y_{\rm s}(u_p)$ the sputtering yield for 
species $s$ integrated over all impact angles and evaluated for an impact velocity of $u_p$. Then the change in rate of the grain radius 
can be determined using the mass of the sputtered species $m_s$ and is given by 
\begin{equation}\label{eq:sputtering}
	\frac{da}{dt} = - \sum_{s,p}{\frac{m_s n_p}{2 \rho_g}  u_p Y_{\rm s}(u_p)}.
\end{equation}
Note that $da/dt$ does not explicitly depend on the grain radius, but it does have a weak dependency 
through the sputtering yield. This is because the relative speed between the impinging neutral or ion species and 
the grains is grain radius-dependent \citep[e.g.][]{VanLooetal2009} and, for small grains, the projectile particle 
may be able to pass through the grain, therefore reducing the sputtering yield at these sizes \citep{bocchio14}.

When $da/dt$ (or $\dot{a}$) is a constant the time evolution of the grain distribution simply reduces to 
\begin{equation}
\frac{\partial n(a,t+\Delta t)}{\partial a} = \frac{\partial n(a-\dot{a}\Delta t, t)}{\partial a},
\end{equation}
and it is possible to split the effect of number-conserving processes from Eq.~\ref{eq:advection}. Then the number density distribution at time $t+\Delta t$ for bin $i$ is given by 
\begin{equation}
	n_i(t+\Delta t) = \int\limits_{a_i}^{a_{i+1}}{\left.\frac{\partial{n(a, t+\Delta t)}}{\partial{a}}\right|_i da}
	= \sum_{j=0}^{N-1}{\int\limits_{a^j_i}^{a^{j+1}_{i+1}}{\left.\frac{\partial{n(a-\dot{a}\Delta t, t)}}{\partial{a}}\right|_j da}}, 
\end{equation}
where $a^j_{i} = \max[a_i, a_j+\dot{a}\Delta t]$ and $a^{j+1}_{i+1} = \min[a_{i+1}, a_{j+1} + \dot{a}\Delta t]$. 
Thus, to determine the evolved number density in bin $i$, it is only necessary to determine from which size bin $j$ the dust grains now residing in bin $i$ come from. This can be done simply by calculating the position of the edges of bin $j$ at time $t+\Delta t$, i.e. $[a_j + \dot{a} \Delta t, a_{j+1} + \dot{a} \Delta t]$, and establishing which bins overlap with bin $i$. The contribution to the number density is then worked out analytically using $A_j$ and $\alpha_j$ describing the power-law distribution in bin $j$ at time $t$. Similarly, we can determine the updated volume density using
\begin{equation}
	\begin{aligned}
		\rho_i(t+\Delta t) &= \int\limits_{a_i}^{a_{i+1}}{m(a)\left.\frac{\partial{n(a, t+\Delta t)}}{\partial{a}}\right|_i da},\\ 
    &= \sum_{j=0}^{N-1}{\int\limits_{a^j_i}^{a^{j+1}_{i+1}}{m(a)\left.\frac{\partial{n(a-\dot{a}\Delta t, t)}}{\partial{a}}\right|_j da}}. 
	\end{aligned}
\end{equation}
From the updated $n_i(t+\Delta t)$ and $\rho_i(t+\Delta t)$ we can solve for the power-law coefficient $A_i$ and the index $\alpha_i$ at time $t+\Delta t$ to find the discrete distribution function $\displaystyle \left.\frac{\partial n(a, t+\Delta t)}{\partial a}\right|_i$.

\subsubsection{Mass-conserving processes}\label{sect:masscons}
In contrast to sputtering, shattering due to grain-grain collisions conserves the total mass of the grain 
distribution but not their total number. Above a threshold impact velocity, some volume fraction of the 
grains involved will fragment into smaller dust grains which themselves follow a size distribution, that is 
$\displaystyle \frac{\partial N_{\rm frag}}{\partial a} \propto a^{-3.3}$ \citep{jones96}.
The evolution of the grain distribution is then described as \citep[e.g.][]{jones94}
\begin{equation}\label{eq:nshattering}
	\begin{aligned}
		S(a,t) &= -\frac{\partial n}{\partial a} \int\limits_{a_{\rm min}}^{a_{\rm max}}{da_1 \frac{\partial n}{\partial a_1} \alpha(a, a_1)}\\
		       &+ \frac{1}{2}\int\limits_{a_{\rm min}}^{a_{\rm max}}{da_1 \frac{\partial n}{\partial a_1}\int\limits_{a_{\rm min}}^{a_{\rm max}}{da_2
	\frac{\partial n}{\partial a_2} \alpha(a_1, a_2) \frac{\partial N_{\rm frag}}{\partial a}(a, a_1, a_2)}},
\end{aligned}
\end{equation}
where $\alpha(a_1,a_2) = \pi(a_1 + a_2)^2 v_{\rm rel}(a_1,a_2)$, when multiplied by the grain number density is 
the collision frequency of grains with radius $a_1$ and
$a_2$ above a threshold velocity for shattering and, otherwise, is equal to zero. 
$\frac{\partial N_{\rm frag}(a, a_1, a_2) }{\partial a} da$ is the number of grains with radii in the range
$[a,a+da]$ produced by interactions of grains with radius $a_1$ and $a_2$ . Note that the first term describes
the removal of dust grains from the interval and the second term the contribution to it due to fragmentation
and requires integration over the entire grain-size distribution.

For the purpose of evolving a discrete distribution,  Eq.~\ref{eq:nshattering} needs to be integrated over the 
different bins. Hence the change of the number density as a function of time for bin $i$ is given by 
\begin{equation}\label{eq:ndisc}
	\begin{aligned}
		S_i(t) =& \int\limits_{a_i}^{a_{i+1}}{da S(a,t)}\\
		=& -\sum_{j=0}^{N-1}{\pi v_{\rm rel}(\langle a\rangle_i,\langle a\rangle_j)\ n_i n_j
	\left\{\langle a^2\rangle_i + 2\langle a\rangle_i\langle a\rangle_j + \langle a^2\rangle_j\right\} }\\ 
	&+ \frac{1}{2}\sum_{j=0}^{N-1}{\sum_{k=0}^{N-1}{\pi v_{\rm rel}(\langle a\rangle_j,\langle a\rangle_k)\ n_j n_k N_{{\rm frag},i}^{j,k}}}\\
	&\ \ \ \quad \times \left\{\langle a^2\rangle_j + 2\langle a\rangle_j\langle a\rangle_k + \langle a^2\rangle_k\right\} ,
\end{aligned}
\end{equation}
where 
\begin{equation}
	\langle a^l \rangle_i = \frac{1}{n_i}\int\limits_{a_i}^{a_{i+1}}{da\ a^l \frac{\partial n}{\partial a}}, 
\end{equation}
where $l$ is an integer, and 
\begin{equation}
	N_{{\rm frag},i}^{j,k} 
			 = \int\limits_{a_i}^{a_{i+1}}{da \frac{\partial N_{\rm frag}}{\partial a}
				 (a, \langle a\rangle_j, \langle a\rangle_k)},
\end{equation}
is the number of grains with sizes between $[a_i, a_{i+1}]$ due to fragmentation by collisions of grains within bin $j$ and $k$. Here, we assumed that the distribution of grain fragments is the same for all grains in bin $j$ and $k$, i.e. 
$\frac{\partial N_{\rm frag}}{\partial a} (a, a_1, a_2) = \frac{\partial N_{\rm frag}}{\partial a} (a, 
\langle a\rangle_j, \langle a\rangle_k)$. If we know the analytic form of the size distribution of fragments and its dependency on the projectile and target radii, a more accurate version of Eq.~\ref{eq:ndisc} can be derived. Furthermore, it is presumed that all grains in a size bin have the same velocity and, thus, the relative velocity between two bins is also constant. Using Eq.~\ref{eq:ndisc}, the number density in bin $i$ at time $t+\Delta t$ is then
\begin{equation}
 n_i(t+\Delta t) = n_i(t) + S_i(t) \Delta t.
\end{equation}
Likewise, the volume density can be updated using
\begin{equation}
	\rho_i(t+\Delta t) = \rho_i(t) + S'_i(t) \Delta t,
\end{equation}
where $S'_i(t)$ can be derived from multiplying Eq.~\ref{eq:nshattering} with $m(a)$ and then discretising the 
integrals. We then find
\begin{equation}\label{eq:rhodisc}
    \begin{aligned}
	    S'_i(t) =& -\frac{4\pi \rho_g}{3}\langle a^3\rangle_i \sum_{j=0}^{N-1}{\pi v_{\rm rel}(\langle a\rangle_i,\langle a\rangle_j)\ n_i n_j}\\
	      & \ \ \ \quad \times \left\{\langle a^2\rangle_i + 2\langle a\rangle_i\langle a\rangle_j +  
         \langle a^2\rangle_j\right\} \\ 
	 &+ \frac{1}{2}\sum_{j=0}^{N-1}{\sum_{k=0}^{N-1}{\pi v_{\rm rel}(\langle a\rangle_j,\langle a\rangle_k)\ n_j n_k m_{{\rm frag},i}^{j,k}}}\\
       &\ \ \ \quad \times \left\{\langle a^2\rangle_j + 2\langle a\rangle_j\langle a\rangle_k + \langle a^2\rangle_k\right\} .
\end{aligned}
\end{equation}
Here the mass transferred to bin $i$ due to collisions of grains within bins $j$ and $k$ is given by
\begin{equation}
	m_{{\rm frag},i}^{j,k} = \int\limits_{a_i}^{a_{i+1}}{da\ m(a) 
\frac{\partial N_{\rm frag}}{\partial a}(a, \langle a\rangle_j, \langle a\rangle_k)}.
\end{equation}
This means that the radius of the fragmented grains is not taken into account, but only an appropriate mass for
all grains within a bin is assumed. Note that this assumption must also be reflected in the mass-loss term, i.e. 
the first term on the left-hand side, as otherwise a systematic discrepancy arises between the mass loss due 
to fragmentation and the redistributed mass across the distribution. Again, such simplifications are not needed
when we know the analytic expression of the fragment distribution in terms of the radii of the colliding grains.
Equations~\ref{eq:ndisc} and \ref{eq:rhodisc} are analogous to the expressions of other authors who have used
either a piecewise-constant or linear description for the discrete distribution function 
\citep[e.g.][McK18]{mizuno88, jones94, jones96, hirashita09}.

\section{Tests and Results}\label{sec:results}
To test the power-law description of the grain distribution we apply the methods of Sect.~\ref{sec:method}
to the test problems outlined in McK18. As these tests have analytical solutions, this allows a direct analysis 
of the performance of the method, but also a direct comparison with both the piecewise-constant and linear 
methods studied in McK18. Note that these tests do not necessarily represent physical or realistic 
situations.

\subsection{Sputtering of a boxcar distribution}\label{sec:boxcar}
Here we will test the convergence of the error in the total grain mass depending on the number of size bins used.
McK18 show that the piecewise-linear method exhibits a $1/N^2$ scaling of the convergence and, thus, second-order
behaviour, which is an improvement of the piecewise-constant method that is only first order.

The initial distribution is taken to be a boxcar function 
\begin{equation}
\setlength{\jot}{10pt}
\frac{\partial n}{\partial a} = \begin{cases}
	1\ {\rm cm^{-4}}&  \text{if} \hspace{4pt}  a_L \leq a \leq a_R\\
	0\ {\rm cm^{-4}}& \text{otherwise.}
\end{cases}
\end{equation}
where $[a_L, a_R] = \left[a_{\rm min}\left(a_{\rm max}/a_{\rm min}\right)^{3/8}, a_{\rm min}\left(a_{\rm max}/a_{\rm min}\right)^{1/2}\right]$ and $a_{\rm min}$ and $a_{\rm max}$ are set to $0.001 {\rm \mu m}$ and 
$a_{\rm max} = 1 {\rm \mu m}$, respectively. Contrary to McK18, who adopt a grain growth rate, we take a constant grain sputtering rate of $\dot{a} = - 2.4\times10^{-7} \hspace{1pt} {\rm cm \hspace{1pt}  Gyr^{-1}}$ applied 
for a time of $t = 5 \ {\rm Gyr}$ in 100 equal time steps. A constant sputtering rate is used here to ensure that the test is analogous to that of McK18. In reality the sputtering rate is size-dependent via the 
sputtering yield \citep[e.g.][]{bocchio14}. Grains which are sputtered to a size smaller than 
$a_{\rm min}$ are assumed to be too small to participate in further sputtering and are removed from the 
distribution. As sputtering only affects the grain mass, the final distribution is still a boxcar function 
between the limits $[a_L + \dot{a}t, a_R+\dot{a}t]$.  

\begin{figure}
\centering
\includegraphics[width=\columnwidth]{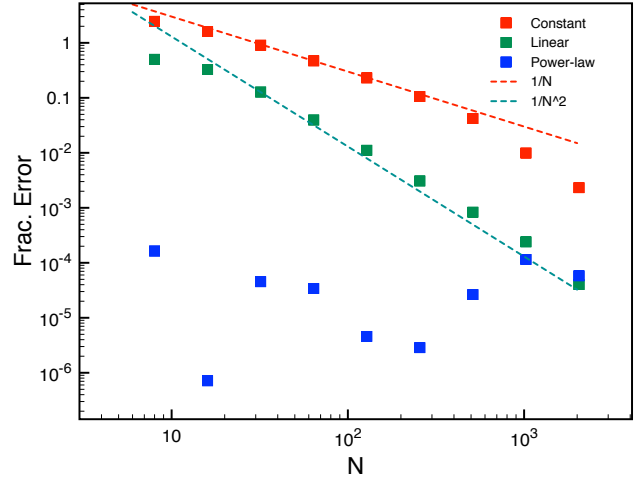}
\caption{Fractional error of the total grain mass as functions of number of bins, $N$, 
	for an initial boxcar distribution affected by sputtering. 
	The boxes show the results for the power-law (blue), piecewise-linear (green)
	and piecewise-constant method (red). The dashed lines show  a $1/N$ scaling (red) 
	and $1/N^{2}$ scaling (green).} 
\label{fig:convergeTest}
\end{figure}

Figure \ref{fig:convergeTest} shows the fractional error in the total grain mass as a function of bin number
(from $N = 8$ to $N = 2048$) for the piecewise-constant, piecewise-linear and power-law methods. Both the 
piecewise-constant and linear methods show their expected first and second order dependence, respectively, 
on bin size and the latter method outperforms the former. However, the power-law method surpasses both
of these with an accuracy below 0.1\% over all bin numbers ($N = 8- 2048$). Especially for a small number of 
bins we find that the accuracy of the power-law method is more than 4 orders of magnitude better than 
the two other methods. The linear method only reaches this accuracy for $N= 512$ and the piecewise-constant 
for $N = 2048$.

\begin{figure}
\centering
\includegraphics[width=\columnwidth]{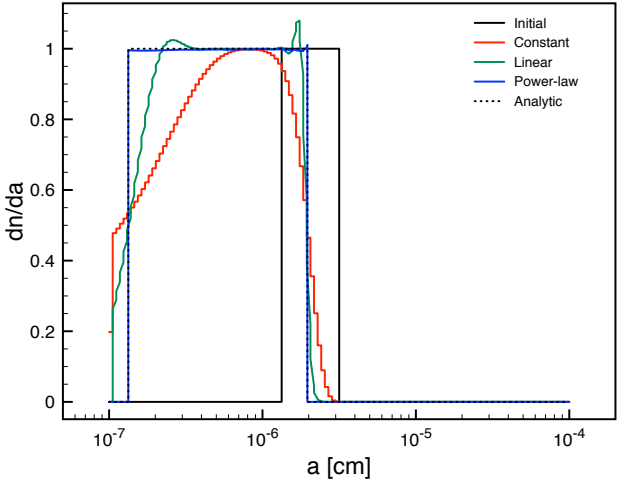}
\caption{Initial boxcar distribution (solid black) with $N = 128$ evolved by applying a sputtering rate of $\dot{a} = - 2.4\times10^{-7} 
	{\rm cm \ Gyr^{-1}}$ for $5 {\rm Gyr}$ using the piecewise-constant (red), piecewise-linear
(green) and power-law method (blue) and compared to the analytic distribution (black dotted).}
\label{fig:boxcardist}
\end{figure}

It is pertinent to understand where these differences in the fractional error between the methods 
come from. In principle all methods should describe the distribution equally well as, for example, 
the piecewise-linear method should reduce to the piecewise-constant method (see Sect. \ref{sec:discrete}). Furthermore, for the power-law method, the power-law index is set to $\alpha = 0$ for $N=2048$ as the root-finding algorithm breaks down (see Sect.~\ref{sec:discrete}). This implies that it also reduces to the piecewise-constant method, yet it produces a result that is nearly two orders of magnitude better than the piecewise-constant method.
The only difference is the treatment of the distribution edges.  As the distribution evolves due to sputtering, it moves across bins, but does not necessarily continue to cover an entire bin at the distribution limits. However, the piecewise-constant method dictates that the grains are uniformly distributed in a bin and, likewise, the linear method uses slope-limiting to distribute the grains across the entire bin. This causes the edge of the discrete distribution to be smeared out at its edges (see Fig.~\ref{fig:boxcardist}). Only the power-law method follows the distribution edges and takes 
them into account when determining the distribution function in the bin. Modifying the piecewise-constant and linear methods to follow the distribution limits, as in the power-law method, leads to an improved accuracy, with the relative errors in the mass below $10^{-4}$ for all bin sizes. Note that the treatment of the distribution edges in the power-law method also produces the variations seen in the relative error as a function of the bin number.
	
\subsection{Sputtering of a MRN distribution}\label{sec:MRN}
While the boxcar distribution of the previous section shows that it is important to carefully treat the edges of the distribution, it is not representative of realistic grain-size distributions. In the ISM the size distribution for silicate and carbon grains is given by a power-law \citep[see Eq.~\ref{eq:MRN}][]{MRN77}. Therefore, in this test, the three methods are tested on their ability to follow the evolution of a power-law distribution affected by sputtering.

We initialise each bin between $[a_{\rm min}, a_{\rm max}]$ with the number and mass density calculated 
using $\displaystyle \left.\frac{\partial{n(a)}}{\partial{a}}\right|_i =  a^{-3.5}$.
We assume the same $a_{\rm min}$, $a_{\rm max}$, sputtering rate and evolution time as in the previous section. 
While we already minimise the errors occurring at the distribution edges by completely filling the full grain-size 
range, we further use the modified piecewise-constant and linear methods as described in the previous section (that is, 
we track the distribution limits). As a result, the distribution is not affected by the issues arising when the edge of the 
distribution falls within a bin, as in the boxcar test, and all the differences are due to the ability of 
each method to describe the underlying grain-size distribution.
\begin{figure}
\centering
\includegraphics[width=\columnwidth]{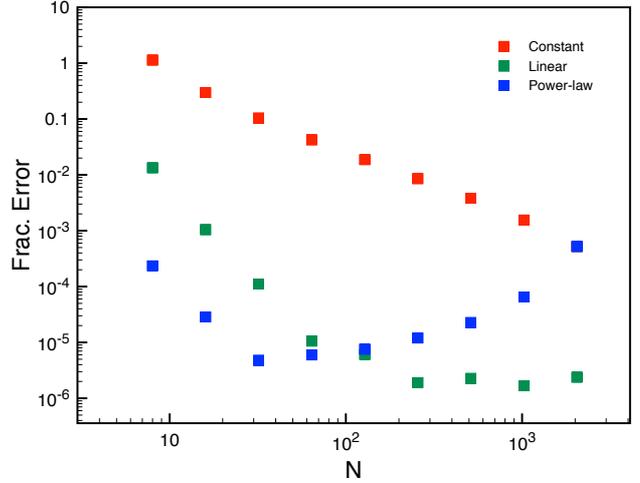}
\caption{Fractional error of the numerical grain mass as functions of number of bins, $N$,
	for an initial MRN distribution affected by sputtering.
	The squares show the results for the power-law (blue) and the modified piecewise-linear (green)
        and piecewise-constant (red) methods. }
\label{fig:MRNconvergeTest}
\end{figure}

Figure~\ref{fig:MRNconvergeTest} shows the fractional error in the total grain mass at the final time for the different methods. While, for the boxcar distribution, the modified piecewise-constant and linear methods have relative errors of the order $10^{-5}$ for all bin sizes, this is no longer true. Especially, the modified piecewise-constant method shows a linear behaviour in the fractional error with large errors at small bin numbers, i.e. $> 10\%$ for $N \leq 32$. The modified piecewise-linear method is significantly better but still performs poorly at small bin numbers, i.e. $N < 16$. Only the power-law method consistently produces errors smaller than $10^{-3}$ for all bin sizes. However, note that, for large values of $N$, the modified piecewise-linear method is better than the power-law method as the latter reduces to the modified piecewise-constant method when the bin size becomes small (see Sect.~\ref{sec:discrete}).
\begin{figure}
\centering
\includegraphics[width=\columnwidth]{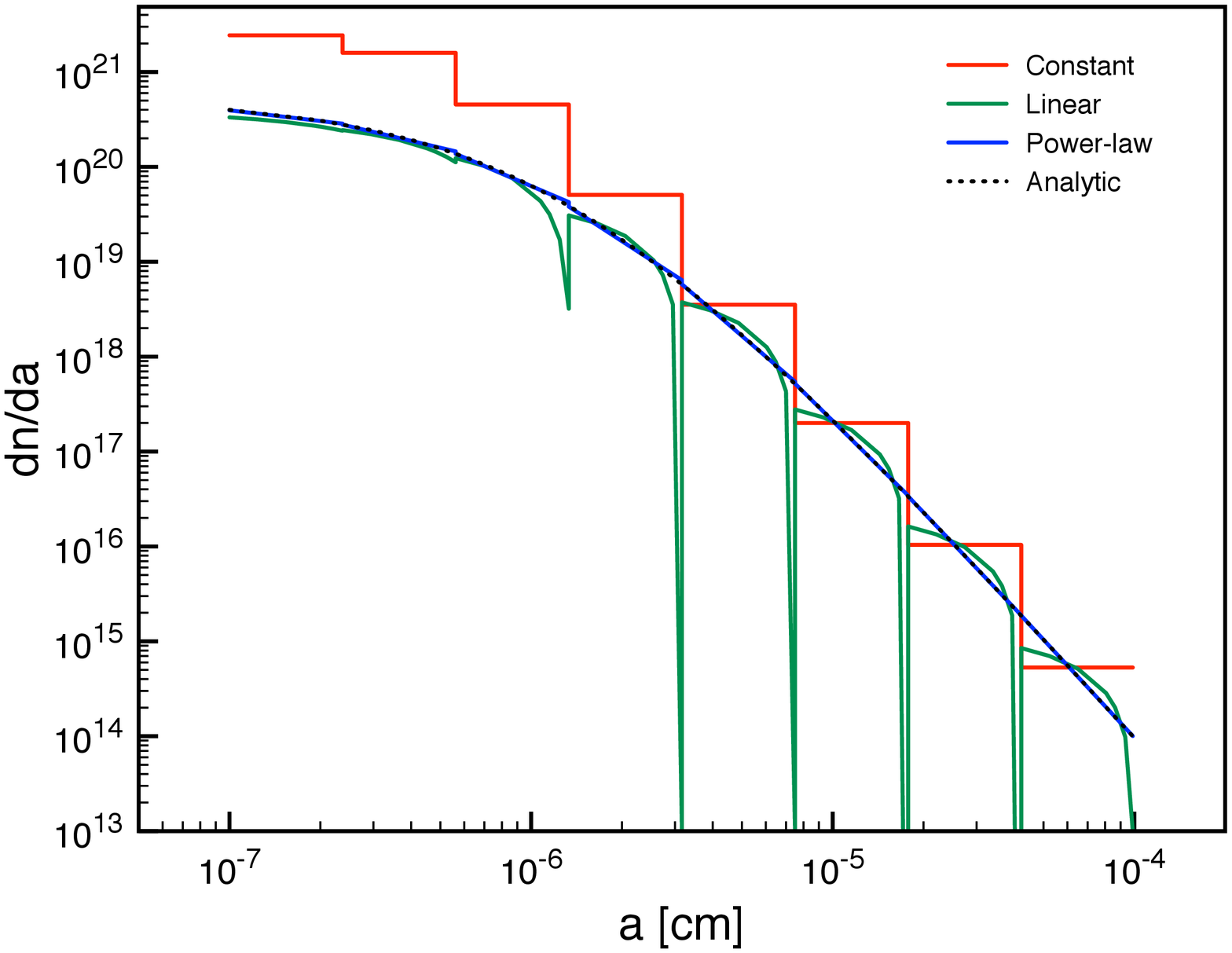}
\includegraphics[width=\columnwidth]{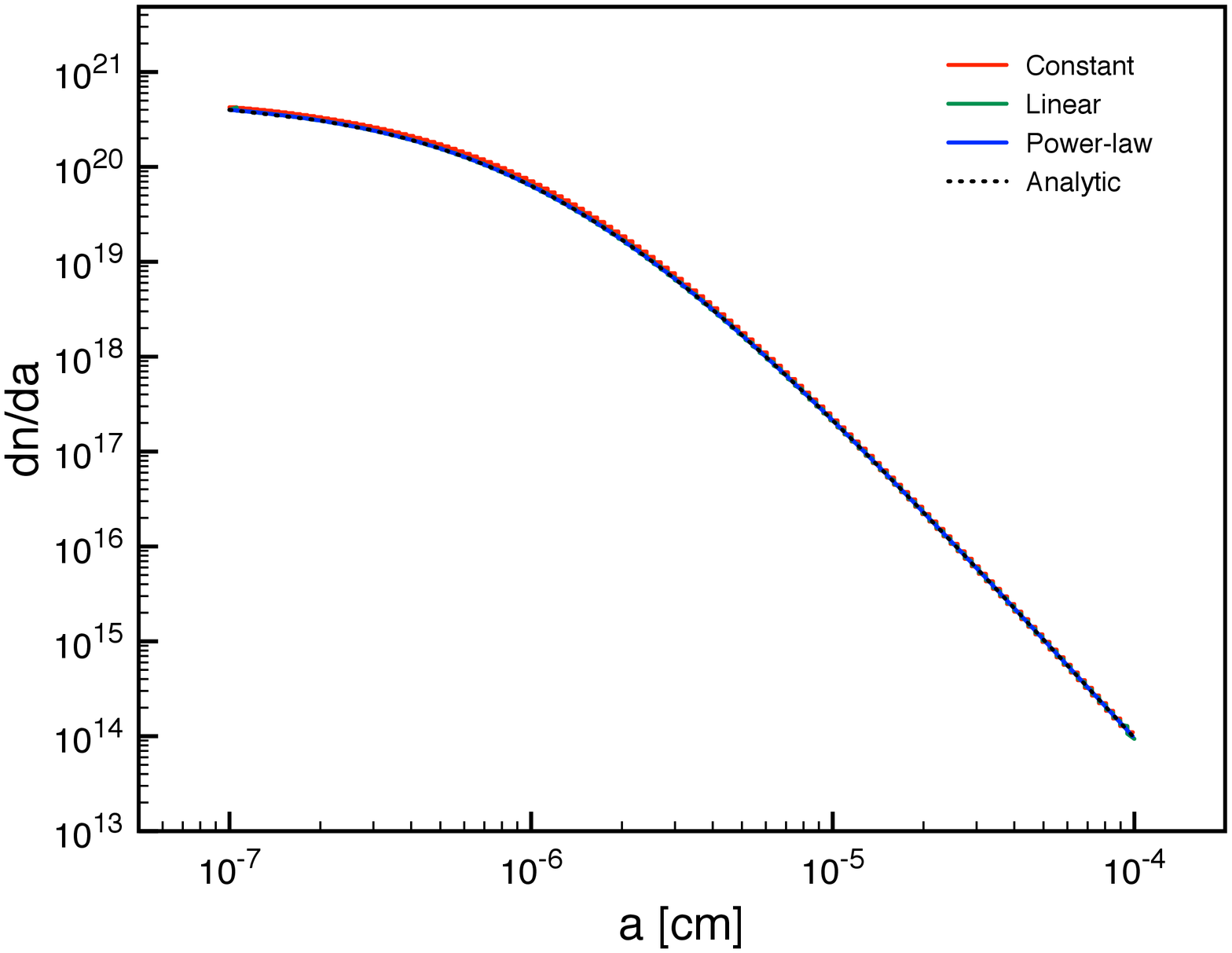}
\caption{MRN distribution evolved by applying a sputtering rate of $\dot{a} = - 2.4\times10^{-7} 
	{\rm cm \ Gyr^{-1}}$ for $5 {\rm Gyr}$ using the modified piecewise-constant (red) and piecewise-linear
       (green) methods and the power-law method (blue) and compared to the analytic distribution (black dotted) for $N=8$ (top)
       and $N=128$ (bottom) bins.}
\label{fig:MRNdist}
\end{figure}

The discrete distribution function for the evolved MRN distribution is shown in Fig.~\ref{fig:MRNdist} for 
$N = 8$ and $128$.  From the figure it is clear that the power-law method describes the power-law distribution 
very well. The modified linear method does capture the analytic solution well at the small grain radii, but
less so at the large radii where it needs to apply slope limiting. However, the modified linear method 
is always better than the modified 
piecewise-constant one when describing a power-law distribution and, as $N$ increases, the 
modified piecewise-linear method converges to the power-law method. Eventually, 
the modified piecewise-constant method will also converge but only at much larger values of $N$. 
This is expected as a power-law distribution, such as in Eq.~\ref{eq:sizeDist}, can be approximated to second 
order as 
\begin{equation}\label{eq:lin2}
	\left.\frac{\partial n(a)}{\partial a}\right|_i = A_i a_0^{-\alpha_i} - \alpha_i A_i 
	a_0^{-\alpha_i - 1} (a - a_0),
\end{equation}
where $a_0$ is a grain size in the interval $[a_i,a_{i+1}]$. The piecewise-linear and piecewise-constant
discretisations are expressed similarly and, thus, eventually converge as the bin size decreases.
Note that the convergence is quicker for shallower power-laws.
\begin{figure}
\centering
\includegraphics[width=\columnwidth]{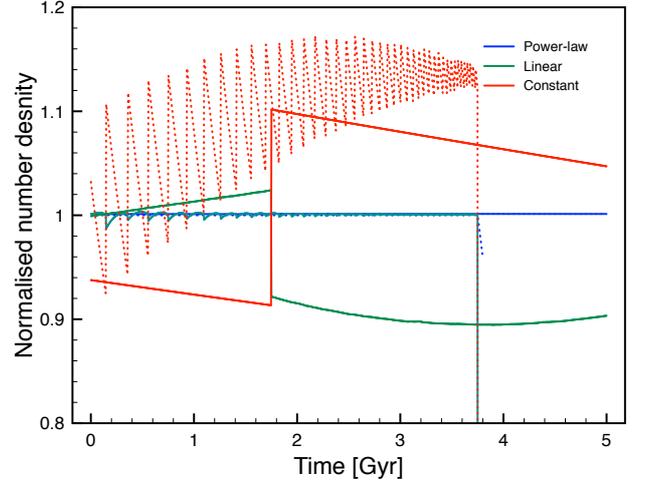}
\caption{Evolution of the number density as function of time for grains with a initial radius of 
        0.01$\rm \mu m$ (dotted) and 0.5 $\rm \mu m$ (solid) for the modified piecewise-constant (red) and
	piecewise-linear (green) methods and the power-law (blue) distributions with $N = 128$. The number
	densities are normalised to the analytic values.  }
\label{fig:diffdist}
\end{figure}

Although all methods are able to conserve the total grain mass and reproduce the final distribution to a 
high degree for $N = 128$ (see Figs.~\ref{fig:MRNconvergeTest} and \ref{fig:MRNdist}), it is also useful to 
evaluate the distribution function at specific grain radii. For processes such as sputtering, as the grains shrink, the number of grains does not change. Figure~\ref{fig:diffdist} shows the grain number normalised to their 
initial value for grains with an initial grain radius of $0.01$ or $0.5 {\rm \mu m}$ for the three methods over a time range of 5Gyr. While the power-law method maintains a constant grain number for both grain radii, both the modified piecewise-constant and linear methods show errors of the order of 10-15\%. These errors do not remain constant but vary significantly over time with large discontinuities when the grains move from one bin to another. Note that the $0.01{\rm \mu m}$ grains move through many more bins than the $0.5 {\rm \mu m}$ ones before they reach $a_{\rm min}$ and are removed from the model 
around 3.8Gyr. Thus, the power-law model does not only preserve global properties of the distribution, but 
also the inter-bin characteristics, unlike the modified piecewise-constant and linear methods.

\subsection{Grain Shattering}\label{subsec:shatter}
The previous tests dealt with grain sputtering, a process which conserves the total number of grains in 
the distribution whilst the total mass of grains in the distribution changes. Here we look at grain shattering, 
in which the total mass of grains is conserved but the number of grains is altered significantly due to the 
production of many small fragments. When two grains of differing sizes, and different velocities, collide 
at a relative velocity exceeding a threshold value, some portion of the grains are fragmented. These smaller 
fragments can themselves be treated as spherical grains that follow a power-law grain-size distribution.

Here we carry out the same shattering test as in McK18. Only the collision between large grains 
($\geq 0.1{\rm \mu m}$) causes fragmentation, with both grains completely destroyed, and, for simplicity,  
the fragments are distributed across the full size range of $[a_{\rm min}, a_{\rm max}]$, where 
$a_{\rm min} = 0.001{\rm \mu \hspace{1pt} m}$ and $a_{\rm max} = 1 {\rm \mu \hspace{1pt} m}$, following a size-distribution 
$\propto a^{-3.3}$. Note that, as fragments can be larger than the fragmenting grains, this model does not consider
only shattering but also some degree of grain growth. The collision velocity between the grains is set to $3~{\rm km s^{-1}}$. 
With these assumptions we have
\begin{equation}\label{eq:nfrag}
	\frac{\partial N_{\rm frag}}{\partial a}(a, a_1, a_2) = \frac{0.7(a_1^3 + a_2^3)}{a_{\rm max}^{0.7} - 
	a_{\rm min}^{0.7}} a^{-3.3},
\end{equation}
resulting in 
\begin{equation}\label{eq:McKnshat}
	N_{{\rm frag},i}^{j,k} = \frac{0.7}{2.3} \left(\langle a^3\rangle_j + \langle a^3\rangle_k\right) 
	\frac{a_i^{-2.3} - a_{i+1}^{-2.3}}{a_{\rm max}^{0.7} - a_{\rm min}^{0.7}},
\end{equation}
and
\begin{equation}\label{eq:McKmshat}
	m_{{\rm frag},i}^{j,k} = \left(\langle m\rangle_j + \langle m\rangle_k\right) 
	\frac{a_{i+1}^{0.7} - a_{i}^{0.7}}{a_{\rm max}^{0.7} - a_{\rm min}^{0.7}}
\end{equation}
where $\langle m\rangle = 4\pi \rho_g/3 \langle a^3\rangle$. Note that the latter expression is the same as 
in McK18 (their Eq.~63) and, to compare results, we will use the same expression. However, as the fragmentation 
distribution, Eq.~\ref{eq:nfrag} has a simple analytic expression, it is possible to derive more accurate, exact 
versions for Eqs.~\ref{eq:ndisc} and \ref{eq:rhodisc}. We will not give those expressions here, but we 
will use those to produce quasi-analytic solutions (using $N=256$ bins) and to evaluate the approximations 
made in $N_{{\rm frag},i}^{j,k}$ and $m_{{\rm frag},i}^{j,k}$. To differentiate between the two power-law methods
we will refer to the former as the default method and the latter as the exact method.
\begin{figure}
\centering
\includegraphics[width=\columnwidth]{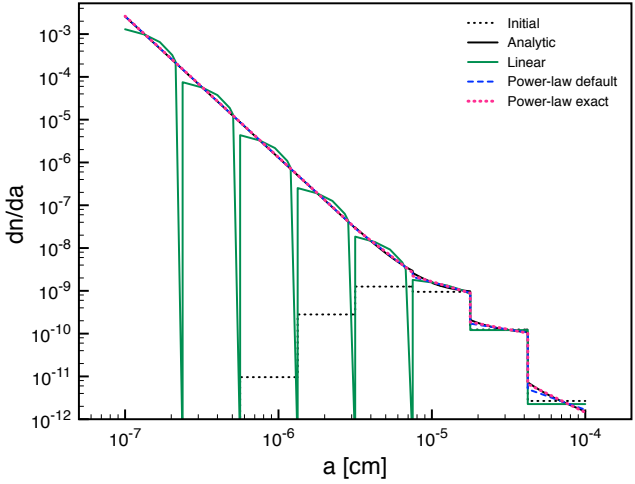}
\caption{Distribution of grains due to shattering after 150 Myr
for the piecewise-linear method (green) and the power-law method. The dashed blue line uses the 
default method with an approximated mass deposited in a bin, while the dotted red line uses the exact method 
using an exact mass calculation. The quasi-analytic solution 
is given by the black solid line and the initial condition by the dotted line.}
\label{fig:shattertest}
\end{figure}
We use the same initial conditions as in McK18, that is a log-normal distribution represented by a piecewise
discrete distribution over $N = 8$ bins and assume that only the largest grains contribute to the shattering as 
$v_{\rm rel}(a_i, a_j) = 3~{\rm km s^{-1}}$ if $a_i \geq 0.1{\rm \mu m}$ and $a_j \geq 0.1{\rm \mu m}$ and is
equal to zero otherwise. Figure~\ref{fig:shattertest} shows the grain distribution due to shattering for 
a time of 150 Myr.  At this time a reasonable amount of large grains have been shattered so that, 
while the initial distribution is only slightly modified at the large radii end of the distribution, the small 
grains follow the $a^{-3.3}$ distribution resulting from the fragmentation. The piecewise-linear method of 
McK18 describes this distribution reasonably well, especially if we evaluate the distribution at 
the geometric midpoint of the bins. Furthermore, quantitatively, the piecewise-linear 
routine produces a relative error in the total number density of about 10\% and conserves the total mass density exactly. However, a closer inspection of the distribution shows that the distribution is not adequately described, particularly at the bin edges. This is due to the slope limiting which needed to be performed at the small grain sizes ($< 0.1 {\rm \mu m}$) in order to ensure positivity of the distribution (while conserving mass). At the same time we see that the distribution of the large grains remains uniform within the bin (that is, a zero slope). This is because, in the method of McK18, the average grain size does not change if a bin loses mass, but only when it gains mass. While it is not a significant problem for this specific test where shattering is treated alone, reproducing the distribution shape becomes important when number-conserving grain processes are also considered. 

The power-law method does describe the grain distribution across the full range of grain sizes more accurately 
with only minor deviations from the analytic solution at the large grain sizes. The exact method performs 
slightly better than the default one in reproducing the analytic distribution. Both methods have a relative 
error below machine precision for the mass density and below 2\% for the number density. Although the 
exact method does describe the distribution better than the approximate method, the errors are similar 
as a discrete distribution with $N = 8$ cannot adequately model the break in the analytic distribution at
$a = 0.1{\rm \mu m}$. This break does not coincide with a bin edge, but instead falls in the middle of a bin.
The benefits of the exact method become clear if we check the relative error in number density in each bin. The total number density is dominated by a single bin with the smallest grain radius, thus errors at the larger grain
radii are not quantified by the relative error in the total grain number density. In the bin with the largest 
radii the relative error is below 0.1\% for the exact method, 8\% for the approximate method and reaches 
25\% for the linear method. The accuracy at which the exact power-law method can reproduce the distribution 
in a bin reflects in improved performance at longer evolution times. Running the shattering test for 
a longer time, for example up to $t = 1$Gyr,  the relative error in the total number density increases to $\approx 10\%$ for the default method, but only to $\approx 5\%$ for the exact method. Thus, it is crucial that the shattering integrals include as much information as possible to minimise the effect of error on the redistribution of fragments across the grain sizes, especially if modelling 
both mass and number-conserving processes.

\subsection{Combined grain sputtering and shattering}
Whilst Sects. \ref{sec:MRN} and \ref{subsec:shatter} show that the power-law method 
performs well for number-conserving processes and mass-conserving processes individually, these processes 
often arise in combination. Therefore, we study here the combined effect of sputtering and shattering 
of grains on an initial MRN distribution.

In this test we will model the dust grain evolution as it occurs within a C-type shock front moving through 
a medium of $n_{\rm H} = 10^6~{\rm cm^{-3}}$ with a dust-to-gas ratio of 0.01. In this situation the dynamics 
of the grains is determined by the balance of 
Lorentz forces and collisions forces with neutral particles. This  results in an effective velocity 
difference between small grains that are coupled to the magnetic field and move with ions and electrons 
and the large grains moving with speeds close to that of the neutrals. \citet{guillet07} 
show that the grain radius at which this transition occurs is between $\sim 7.5\times 10^{-6}$ and 
$2.5\times 10^{-5}$ cm depending on the density of the gas. Here we assume a discontinuous transition at
\begin{equation}
	a_{t} = a_{\rm min}\left(\frac{a_{\rm max}}{a_{\rm min}}\right)^{\frac{3}{4}} \approx 1.17\times10^{-7}{\rm cm },
\end{equation}
where $a_{\rm min}$ and $a_{\rm max}$ are the same distribution limits as used in all previous tests, 
and that the velocity difference is $15~{\rm km s^{-1}}$. (Note that the transition is always on a bin edge.)
Hence, only the small grains, $a < a_t$, will experience non-thermal sputtering due to neutral species, while shattering is 
due to collisions of small grains with large grains. For simplicity, we apply the same shattering procedure 
as in the previous section, that is both grains completely shatter and the fragments are distributed across 
the full range of grain sizes.  Using Eq.~\ref{eq:sputtering} we can estimate the
rate at which the grain radius decreases, i.e. we find $\frac{da}{dt}\approx -10^{-12}~{\rm cm~s^{-1}}$. 
We evolve the distribution for $10^6$s. As there is no analytic solution for this problem, we assess the 
results using the converged solution for the distribution function as the bin number increases. We find that 
both  the linear and power-law methods converge to the same solution. 

\begin{figure}
\centering
\includegraphics[width = \columnwidth]{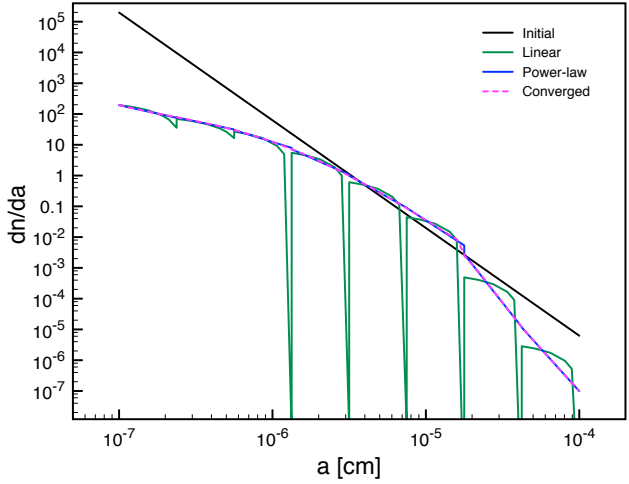}
\caption{Final distribution after sputtering and shattering are applied in combination for a 
time of $10^{6}$s with the sputtering rate $-10^{-12}$ cm ${\rm s^{-1}}$ for the piecewise-linear 
(green) and power-law (blue) methods ($N =8$). The initial MRN distribution is also shown as 
well as the converged solution (dashed magenta, $N = 256$).}
\label{fig:spShRate1}
\end{figure}

\begin{figure}
\centering
\includegraphics[width = \columnwidth]{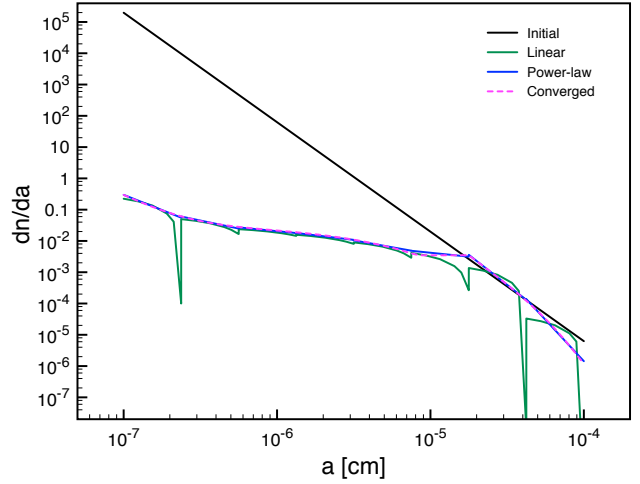}
\includegraphics[width = \columnwidth]{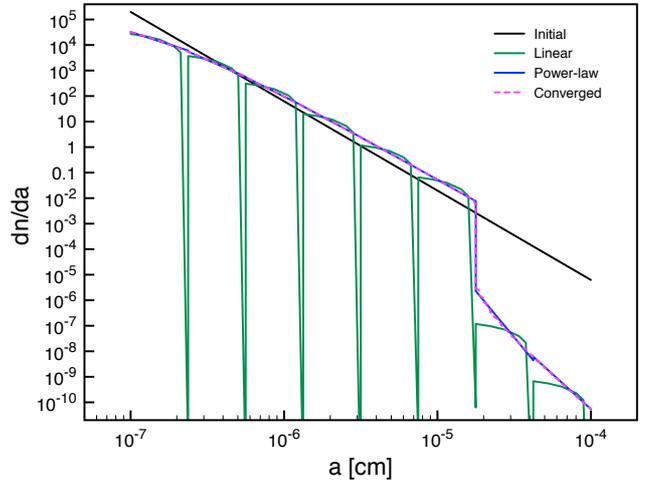}
\caption{Same as Fig. \ref{fig:spShRate1} but with a sputtering rate of $-10^{-11}$~cm~${\rm s^{-1}}$
(top) and $-10^{-13}$~cm~${\rm s^{-1}}$ (bottom).}
\label{fig:spShRate2}
\end{figure}

Figure \ref{fig:spShRate1} shows the grain distribution for the piecewise-linear and the power-law 
methods using $N=8$ bins. Comparing the results with the initial MRN distribution we find that sputtering
changes the slope of the distribution towards the small grains while shattering changes it for the large grains, 
an effect previously noted by e.g. \citet{bocchio14, bocchio16, kirchschlager19}.
Both the power-law and piecewise-linear methods are close to the converged distribution function, although 
the linear method is affected by slope limiting to ensure positivity of the distribution function. Slope 
limiting conserves mass, but not numbers, and this is reflected in the error relative to the converged
distribution function. The linear method has a relative error in the total mass of only 2\% but it is 7\% in 
the total numbers. In comparison, the relative errors for the power-law method are below 1\% even for $N=8$ bins.
However, the piecewise-linear method does converge quickly and achieves the same accuracy with $N=32$ bins.

The results of this test depend on the relative strength of the sputtering and the shattering. Therefore,
as the linear and power-law methods perform differently for number-conserving and mass-conserving processes,  
we also perform the test with a sputtering rate an order of magnitude larger and smaller. 
Figure~\ref{fig:spShRate2} shows the grain distributions for these two additional models. For the higher
sputtering rate, we find that the evolution of the distribution is dominated by sputtering. The sputtering removes more 
small grains from the distribution compared to the model with the default rate. Hence, less projectiles are
available to shatter the large grains and, consequently, the distribution function at large grain radii does not evolve as much.   
Both the piecewise-linear and power-law methods with $N=8$ bins are close to the converged solution with the relative error in the total mass
about 2\% for the linear method and 0.1\% for the power-law method. However, the error in the 
the total numbers is up to 20\%  for the linear method while it is less than 1\% for the power-law method.
The linear method achieves the same performance as the power-law method for $N=64$ bins. 
For the lower sputtering rate, the evolution is mainly due to shattering as the number of large grains 
drops significantly. Sputtering does not remove many grains at the small radii so more projectile grains 
can collide with the large grains and shatter them.
Again both methods reproduce the converged distribution very well, and this also is revealed
in the relative errors. The relative error in the total grain number for the linear method (which is always
the largest error) is only 3\%.

This test shows that the power-law method maintains its high level of accuracy for $N= 8$ bins even when 
combining number-conserving and mass-conserving processes. To achieve the same level with the piecewise-linear
method one needs to model the distribution with more bins, i.e. $N > 32$. Additionally, as with the previous 
tests, the power-law method is also able to produce the correct shape of the final distribution with only 
$N = 8$ bins, something which the linear method has been unable to achieve across all tests. This further 
enforces the usefulness of the power-law method for following the evolution of a grain-size distribution when 
limited computational resources are required or necessary. 

\section{Implementation in hydrodynamical codes}\label{sec:equations}
Although the focus of this paper is on following the evolution of a grain-size distribution due to grain processing,
our goal is to use this discrete power-law prescription in hydrodynamical simulations. To incorporate the 
grain-size distribution into, for example, a multifluid MHD code such as the one of \citet{vanloo13} 
the equations for the grain fluids need to be altered. 

The starting point for this modification are the dynamical equations for each individual (pressureless) grain fluid
which, in a weakly ionised plasma, are given by the continuity equations and the reduced momentum equation
\begin{equation}\label{eq:MHD}
\begin{aligned}
&\frac{\partial n}{\partial t} + \nabla\cdot \left(n {\bf v}\right) = S(a,t) + S_{\rm sputt}(a,t),\\
&\frac{\partial \rho}{\partial t} + \nabla\cdot \left(\rho {\bf v}\right) = S'(a,t) + S'_{\rm sputt}(a,t),\\
&\alpha \rho \left({\bf E} + {\bf v} \times {\bf B}\right) + \rho \rho_{n} K_{gn} ({\bf v}_n - {\bf v}) = 0,
\end{aligned}
\end{equation}
where $S(a,t)$ is the grain shattering loss term given by Eq.~\ref{eq:nshattering} for
the number density, $S'(a,t)$ the shattering loss term for the mass density, $S_{\rm sputt}(a,t)$ and 
$S'_{\rm sputt}(a,t)$ the corresponding terms for grain sputtering, $\alpha = Ze/m$ the charge-to-mass 
ratio of the grains, $\bf{E}$ and $\bf{B}$ the electric field and magnetic field of the medium, and  $\rho_n$ and ${\bf v}_n$ the neutral mass density and velocity.  Also, $K_{gn}$ is the collision coefficient between the grain and the neutrals given by \citep{draine86}
\begin{equation}\label{eq:Kng}
   K_{gn} = \frac{8}{3} \frac{\pi a^2}{m_n + m}\left(\frac{2k_BT_n}{\pi m_n}\right)^{1/2} 
     \left(1 + \frac{9\pi ({\bf v}_n - {\bf v})^2}{128k_BT_n}\right)^{1/2},
\end{equation}
with $T_n$ the temperature of the neutral gas and $m_n$ the mass of a neutral particle. Since we are modelling grains in a weakly ionised plasma, the collision frequency between grains and charged particles is negligible and we need only 
consider grain-neutral collisions. As $n$ and $\rho$ in Eq.~\ref{eq:MHD} are effectively $\displaystyle \frac{\partial n}{\partial a}da$ and $\displaystyle  m(a) \frac{\partial n}{\partial a}da$, we can find the governing equations by integrating the above equations over the range of radii for a given bin. For bin $i$, these then become
\begin{equation}
\begin{aligned}
&\frac{\partial n_{i}}{\partial t} + \nabla \cdot \left( n_{i} \bar{{\bf v}}\right) = S_i + S_{i,{\rm sputt}},\\
&\frac{\partial \rho_{i}}{\partial t} + \nabla \cdot \left(\rho_{i} \bar{{\bf v}}\right) = S'_i + S'_{i, {\rm sputt}},\\
&\langle Z\rangle_i e n_i \left({\bf E} + \bar{{\bf v}} \times {\bf B}\right) + n_i \rho_{n} K^*_{gn} ({\bf v}_n - \bar{{\bf v}}) = 0,
\end{aligned}
\end{equation}
where $S_i$ and $S'_i$ are given by Eqs~\ref{eq:ndisc} and \ref{eq:rhodisc},  $S_{i,{\rm sputt}}$ and 
$S'_{i,{\rm sputt}}$ represent the sputtering losses in bin $i$ and $\langle Z \rangle_i e$ the average grain 
charge. Furthermore, 
\begin{equation}
 K^*_{gn} = \frac{8}{3} \pi \langle a^2\rangle_i \left(\frac{2k_BT_n}{\pi m_n}\right)^{1/2} 
 \left(1 + \frac{9\pi ({\bf v}_n - \bar{{\bf v}})^2}{128k_BT_n}\right)^{1/2},
\end{equation}
is the mean specific collision coefficient between neutrals and grains in bin $i$.
	
Note that, in order to derive these expressions, the only assumption we have made is that all grains within a bin 
have the same velocity. This is similar to the premise made for the collision frequency in the shattering 
process (see Sect.~\ref{sect:masscons}). However, from the reduced momentum equation, it is clear that the 
grain velocity depends on the grain radius through the Hall parameter (the ratio of the gas-grain collision frequency to the gyrofrequency) $\beta = Ze B/m\rho_n K_{gn} \propto a^{-1}$. For small Hall parameters, that is $\beta < 0.1$, the grains move with the neutrals, while, for $\beta > 2$, they move with the electrons and ions. Thus, there is only a small range of $\beta$, or grain radius, for which grains have a velocity in between and only in the bin where this transition occurs can some error in the dynamics be expected. In a subsequent paper studying grain processing in C-type shocks, we will analyse this further.

To deal with the grain processing due to shattering and sputtering in a hydrodynamical code, the routines
described in Sect.~\ref{sec:method} are used. Shattering can be included as a source function during the advection update, while we have the option to do the same for the sputtering or to operator split. The latter is preferred as the method described in Sect.~\ref{sect:numbcons} already gives the updated grain distribution function. Optimally, the operator split is done using Strang splitting with half a time step before the advection update and half of a time step after it.  

\section{Discussion and Conclusions} \label{sec:conclusions}
In this paper we present a numerical method to follow the evolution of a dust grain-size distribution
undergoing grain processes which either conserve grain mass or grain numbers. Guided by observations
of typical ISM dust distributions, our method uses a power-law prescription to specify the distribution 
within a bin. Using the number and mass density of grains within a bin, the coefficient and index of the power 
law can be uniquely determined. We also explicitly track the grain size limits of the distribution. Furthermore, we describe the methods to evolve the discrete power-law distribution due to number-conserving or mass-conserving grain-processes and illustrate this with grain sputtering and grain shattering. The power-law method is complementary to the methods employing either a discrete piecewise-constant or piecewise-linear distribution \citep[e.g.][]{mizuno88, jones96, mckinnon18}
	
The tests performed here show that the power-law method significantly outperforms both the piecewise-constant
and linear methods for following the evolution of the distribution function, especially when the distribution 
is covered by a small number of bins. The main reason is, of course, that the discrete power-law method is naturally 
suited for modelling a continuous power-law distribution as often occurs in the ISM. The linear and the constant
method only provide a second order and a first order  approximation respectively. In part it is also because we follow the distribution lower and upper limits and take them into account when deriving the distribution properties. This is important when considering number-conserving processes and the full radius range of a bin is not filled. For these processes, both the piecewise-constant and linear methods then diffuse the distribution limits. By implementing the same technique as in the power-law method, the relative errors can be reduced in the other two. 

The power-law method is more effective for treating mass-conserving processes than the other methods,
with the best results occurring when more information of the physical processes (in our case grain shattering) 
is included when evaluating the integrals. All methods conserve mass to machine precision accuracy, but the 
number density of the grains is better reproduced with the power-law method. While uncertainties are 
expected when modelling physical processes, it is best to avoid numerically induced ones. As mass- and number-conserving processes are often modelled together, the combined shattering and sputtering test demonstrated that the power-law method will provide the best results especially for small bin numbers (that is $N=8$). For larger bin numbers both the power-law and piecewise-linear methods produce similar results.

The aim of this work is to provide an efficient numerical method that describes the evolution of a dust grain distribution
due to advection and grain physics accurately in large-scale simulations. To avoid a large demand in numerical resources, it is beneficial to cover the grain distribution with a minimal number of bins. As our power-law method 
produces very small errors even for $N=8$, it is perfect to include this approach in a numerical hydrodynamics 
code. One drawback is that operations such as $\rm pow()$, $\rm \log()$ and $\rm \sinh()$ are considerably more 
CPU expensive than linear operations. This is especially important when finding the root of Eq.~\ref{eq:root}.
Using standard algorithms, the power-law method using $N=8$ is only as fast as the linear method with $N = 128$. 
However, one can use alternative algorithms and approximations for these operations so that the CPU cost of the 
power-law method is only 1.5 times that of the piecewise-linear method for the same number of bins. Thus, the 
power-law method not only provides a more accurate, but also viable, alternative to the piecewise-linear method. 
The implementation of the grain-processing methods in a hydrodynamics code is straightforward and only
needs minimal alterations to the equations for a single dust grain fluid. The main assumption is a constant velocity 
for all the grains within a bin. In a weakly ionised plasma this assumption is likely not to be restrictive. In a subsequent 
paper we will further investigate this assumption when modelling C-type shocks with a full dust grain distribution. 

Although the focus has been on the evolution of dust grain-size distributions, the application of this method is not limited to dust alone. The power-law method can be used to follow the evolution of any distribution function, 
especially those which exhibit a power-law distribution. One possible field of application is, for example, the energy loss
of a cosmic ray distribution due to synchrotron radiation or inverse-Compton scattering.   

\section*{Acknowledgements}
We thank the anonymous referee for their helpful comments which have improved the paper.
We thank Tom Hartquist for the useful discussions on grain physics and Paola Caselli for her continued interest and
support. RS thanks the  STFC  and the Center for Astrochemical Studies at the Max Planck Institute for Extraterrestrial 
Physics for funding this PhD project. SVL is supported by a STFC consolidated grant. The data presented
in this paper is available at  https://doi.org/10.5518/751.



\bibliographystyle{mnras}
\bibliography{refs} 



\appendix

%


\bsp	
\label{lastpage}
\end{document}